# Theoretical study of metal-encapsulating Si cage clusters: Revealing the nature of peculiar geometries

**T Miyazaki[1,2], H Hiura[1,3] and T Kanayama[1,2]**
[1]MIRAI Project, [2]ASRC-AIST, [3]ASET
[1,2,3]AIST Tsukuba Central 4, Higashi 1-1-1, Tsukuba 305-8568, Japan

**Abstract.** We present a density-functional (DF) study of structures of Si cage clusters that encapsulate metal atoms. As a prototypical example, the case of $WSi_n$ clusters is shown. To obtain the low-energy clusters in efficient and unbiased ways, genetic-like geometry updates have been performed for generating inputs for subsequent local optimizations within DF calculations. Well-defined cages occur for a certain range of $n$. Such cages are modelled as simple 3-polytopes where the numbers of their inner diagonals close to the metal atom are maximized.

## 1. Introduction

Synthesis of fullerene-like clusters composed of silicon atoms (Si) is very exciting, because it is a challenge against a well-established thinking that only carbon atoms (C) can stabilize a fullerene cage. The subject is also of a great technological impact. One may expect a wealth of structural, physical, chemical and electronic properties of fullerene-like Si clusters, many of which should be useful in Si-based nano-technology.

Is it in principle possible to stabilize a fullerene-like Si cage? The purpose of the present work is to give an answer to this question from a theoretical point of view. Our conclusion is that it should be possible, by doping a suitable metal atom ($M$) inside the cage. The cage is a simple 3-polytope which maximizes the number of its inner diagonals close to the metal atom.

Roughly speaking, stability of a carbon fullerene cage results from the fact that both σ bonding of $sp^2$ and π interactions among the $p_\perp$ orbitals of C atoms contribute to the total binding, where $p_\perp$ is the $p$ orbital component perpendicular to the cage surface. Formation of the Si counterpart seems unlikely to occur because the latter is missing. This necessitates doping of $M$, which induces an additional strength to the cage by $M$-Si binding due to the interactions of orbitals of $M$ with $p_\perp$ orbitals of Si.

Indeed, Hiura et al.[1] have found some fingerprints of formation of such Si cage clusters although it is still unclear whether the cage structures are fullerene-like. The have grown $MSi_nH_x^+$ clusters from vapor of $M$ ($M$=Hf, Ta, W, Re, Ir etc.) and molecules of silane ($SiH_4$) and found that there is a "magic number" $n_0$ for which the cluster is particularly stable. Strikingly, such stable clusters are almost free from H ($x \approx 0$) despite that the clusters must have experienced an H-rich environment. Since metal atoms are chemically reactive, they should be surrounded by Si atoms and also greatly stabilized with $n=n_0$. An *ab initio* structure optimization of a $WSi_{12}$ cluster has revealed that the W atom favors to be encapsulated in the center of a Si cage with a regular hexagonal prism configuration.

There have been some theoretical studies of metal-atom encapsulated Si and Ge clusters[2-9]. However, it appears that most of them do not make serious efforts to find out lowest possible energy structures of the clusters that should be the premise to study cluster science, but simply present results of local optimizations by starting from a few initial structures. Nor they consider any reason for why the cluster structures they have obtained are energetically favorable. In contrast, we have attempted an extensive search for the global-energy-minimum (GEM) structures of $WSi_n$ clusters. We shall also present here a clear explanation to the origin of peculiar geometry of energetically favorable structures of the clusters that we have found from a topological point of view.

## 2. Method of calculation

The procedure of exploring the GEM structure taken by us is partitioned into two stages: (i) global migration on the potential energy surface (PES) and (ii) local optimization of the candidate structures found in (i). The genetic-like structure updates using the single-parent evolution algorithm (SPEA) proposed by Rata et al.[10] have been applied to performing stage (i), while the quenched molecular dynamics simulation has been done in stage (ii). The SPEA may be regarded as an improved version of a genetic algorithm (GA) implemented and applied to the GEM searches of carbon clusters by Deaven and Ho[11]. In GA, one usually assumes two or more parent clusters to generate children, where the number of parents and the method of mating to generate the children must be tuned. Rata et al. have noticed that the mating operation among parents in the Deaven-Ho GA can be replaced by "piece reflection" and "piece rotation" operations of *one* parent cluster, which is why the new method is called the *single-parent* evolution algorithm. Reduction in the parent number should greatly accelerate the efficiency of global optimization. In stage (i), however, we have been faced with a severe difficulty for its goal: the lack of efficient and accurate parameters to describe the W-Si interactions. Therefore we have had to perform both (i) and (ii) processes at the DF level of theory, which limits the total sampling number of the PES points. Concretely, typically 50 SPEA structure updates have been done per cluster in stage (i). The update number is much smaller than have been taken in optimizations of pure Si clusters (see below). Thus we do not insist at present that the lowest-energy structure of a $WSi_n$ cluster that we show is of the global energy minimum.

Another technical aspect of using SPEA is that an update of a cluster structure is so drastic that it may be often necessary to "massage" the as-updated cluster prior to being used as the input for the subsequent DF calculation.

In the GEM searches of pure Si clusters [12], the simulation usually starts from global sampling of the PES by using tight-binding (TB) parameters whose accuracy to reproduce some quantities of a bulk Si crystal has been established relatively well. This is the stage (i) in our terminology. In general, genetic and/or stochastic structure updates are employed in this TB run with typical 1,000 to 10,000 updates, each of which is further followed by 10 to 100 steps of local structure optimizations by using a conjugate-gradient or molecular-dynamics method. If one feels that the system has been located close to the GEM of PES through the TB run, then one switches the electronic structure calculation to that at the DF level of theory and further local optimization is performed to place the TB-calculated candidate structure at the minimum point. This latter part is the stage (ii) in our classification.

It should be emphasized that, even in this "relatively easy" problem of Si$_n$ clusters with $n$ ranging from 20 to 30, their lowest-energy structures, which the authors in the literatures call the GEM structures, have been "updated" year by year [10]. This means the obvious difficulty to perform *ab initio* searches for *true* GEM structures of Si clusters. One should keep in mind that it is even more difficult to do the same for metal-encapsulated Si clusters because of the lack of efficient W-Si interaction parameters to be used in the stage (i).

Our mission is to find out lowest possible energy structures of WSi$_n$ clusters ($n$=10, 12, 14 and 16). The total energy of a cluster is calculated by using the standard density-functional theory within the generalized gradient approximations (GGA) to the exchange-correlation energy. In stages (i) and (ii), the GAUSSIAN98 [13] and STATE [14] codes have been used, respectively.

## 3. Results

### 3.1. Topology of a "fullerene-like" cage

Prior to presenting results of our DF calculations, we should explain what we mean by "fullerene-like". There have been at least two definitions of a fullerene, the IUPAC [15] and CAS [16] versions. The former is "Fullerenes are defined as polyhedral closed cages made up entirely of *n* three-coordinate carbon atoms and having 12 pentagonal and ($n$/2-10) hexagonal faces, where $n \geq 20$. Other polyhedral closed cages made up entirely of *n* three-coordinate carbon atoms shall be known as *quasi*-fullerenes." In the latter, fullerenes are defined as "the even-numbered, closed spheroidal structures of 20 or more carbon atoms, in which every atom is bonded to three other atoms".

It is possible to define fullerenes in a much simpler and more beautiful way than the above two, in the language of topology. For this purpose, we introduce two concepts developed in topology, a convex simple 3-polytope *P* and an inner diagonal [17]. The former is a polytope in three dimensions where each vertex is incident to precisely three edges. An inner diagonal of *P* is a segment that joins two vertices of *P* and that exists, except for its ends, in *P*'s relative interior. The total number of inner diagonal, $\delta_3$, is an important quantity to characterize *P*. Conventional fullerene (C$_n$, $n \geq 20$) geometries are of convex simple 3-polytopes with *maximum* $\delta_3$. Here we may define a fullerene topology as that of a convex simple 3-polytope with *maximum* $\delta_3$. This generalizes the use of the term fullerene to clusters with $n \leq 20$. For example, we can call a cube and a pentagonal prism fullerenes with eight and ten vertices, respectively.

Let us explain why we can replace the conventional definitions of fullerenes by a new one. We introduce a vector notation ($f_4,f_5,f_6$) to specify the structure of *P*, where $f_k$ is the number of *k*-membered rings of *P*. Any *P* with $k$=3 or $k \geq 7$ is not considered, because possible values of *k* are limited to 4, 5 and 6 for *P* with $n$>4 and maximum $\delta_3$. A famous Euler's theorem for a closed polytope is

$$V + F = E + 2, \quad (1)$$

where *V, F* and *E* are the number of vertices, facets and edges of the polytope. If each vertex is connected to three neighbors, then $E$=3$V$/2, meaning that *V* must be even. Combination of this with Eq.(1) and $F$=$f_4$+$f_5$+$f_6$ leads to

$$V = 2f_4 + 2f_5 + 2f_6 - 4. \quad (2)$$

On the other hand, substitution of $E$=4$f_4$+5$f_5$+6$f_6$ and $F$=$f_4$+$f_5$+$f_6$ into Eq.(1) gives us

$$V = f_4 + \frac{3}{2}f_5 + 2f_6 + 2. \tag{3}$$

Solving Eqs.(2) and (3) for $f_4$ and $f_6$, we obtain

$$(f_4, f_5, f_6) = \left(6 - \frac{f_5}{2}, f_5, \left(\frac{V}{2} - 4\right) - \frac{f_5}{2}\right). \tag{4}$$

An important consequence of Eq.(5) is

$$0 \leq f_5 \leq min(12, V-8), \tag{5}$$

that is, *maximum possible value* of $f_5$ is $V-8$ for $V \leq 18$ and $f_5=12$ for $V \geq 20$. Exceptions occur at $V=18$ and 22, where maximum $f_5$ are 8 and 10, respectively[17].

We now turn to evaluation of $\delta_3$. There are two other diagonals, 1-diagonals (edges) and 2-diagonals (diagonals inside a $k$-membered ring), than inner diagonals (or 3-diagonals). Denoting the number of 1- and 2-diagonals as $\delta_1$ and $\delta_2$, respectively, the following relationship holds among $\delta_1$, $\delta_2$ and $\delta_3$:

$$\delta_1 + \delta_2 + \delta_3 = V(V-1)/2, \tag{6}$$

from which we have

$$\delta_3 = \frac{V(V-13)}{2} + \frac{f_5}{2} + 24, \tag{7}$$

where $\delta_2 = 2f_4 + 5f_5 + 9f_6$ and Eq.(4) are substituted into $f_4$ and $f_6$. In Eq.(7), we see that, at a given $V$, *a maximum $\delta_3$ occurs for a maximum $f_5$*, which is 12 and also $f_4=0$ for $V \geq 20$ except for $V=22$ which is 10 and $f_4=1$. Thus convex simple 3-polytopes with maximum $\delta_3$ precisely meet the conditions of fullerene structures defined in both the IUPAC and CAS versions. It is evident that one can also extend the concept of fullerenes to polytopes of $V$ less than 20. Since the previous definitions of fullerenes are only for clusters of carbon atoms, we call clusters of Si atoms with the same topologies as fullerenes fullerene-like Si cage clusters.

*3.2.    Structures and energies of WSi$_n$ clusters*

Figure 1 presents low-energy structures of WSi$_n$ clusters ($n$=10, 12 and 14). For $n$=10 (Fig.1(a)), the Si atoms do not create a well-defined cage although they surround the W atom. This is probably because there are too few Si atoms to gain substantial energy due to Si-Si interactions. In other words, it appears that there is a minimum number of Si atoms that create a smooth cage around a metal atom. For $n$=12 and 14, fullerene-like cages of Si atoms are generated. The lowest-energy structures of WSi$_{12}$ and WSi$_{14}$ have (6,0,2) (Fig.1(c)) and (3,6,0) (Fig.1(d)) cages, respectively. The $\delta_3$ of the former is 18. Although the structure with the maximum $\delta_3$ is of a (4,4,0) cage, the corresponding total energy is 2.2 eV higher than the (6,0,2) cage. On the other hand, $\delta_3$ of the latter (34) is the largest among the convex simple 3-polytopes with 14 vertices.

In order to make a deeper understanding of these results, we count the number of inner diagonals passing close to the metal atom, which we denote as $\delta_3^{eff}$ (Table 1). Since it depends on the cut-off radius $d_{cut}$ to define the vicinity of the metal atom, we use $\delta_3^{eff}$ as counted with $d_{cut}$ =0.25 A. We find for WSi$_{12}$ that $\delta_3^{eff}$ is 6 for a (6,0,2) cage but zero for a (4,4,0) cage. It is obvious that $\delta_3^{eff}$ approximates the magnitude of the overlap between metal-atom orbitals and Si $p$ counterparts that is crucial to stabilization of a Si cage. We also find for WSi$_{14}$ clusters that $\delta_3^{eff}$ is largest (4) for a (3,6,0) cage. The $\delta_3^{eff}$ is 2 and zero for a (5,2,2) cage and for both (4,4,1) and (6,0,3) cages, respectively. The lower the energy, the larger the $\delta_3^{eff}$.

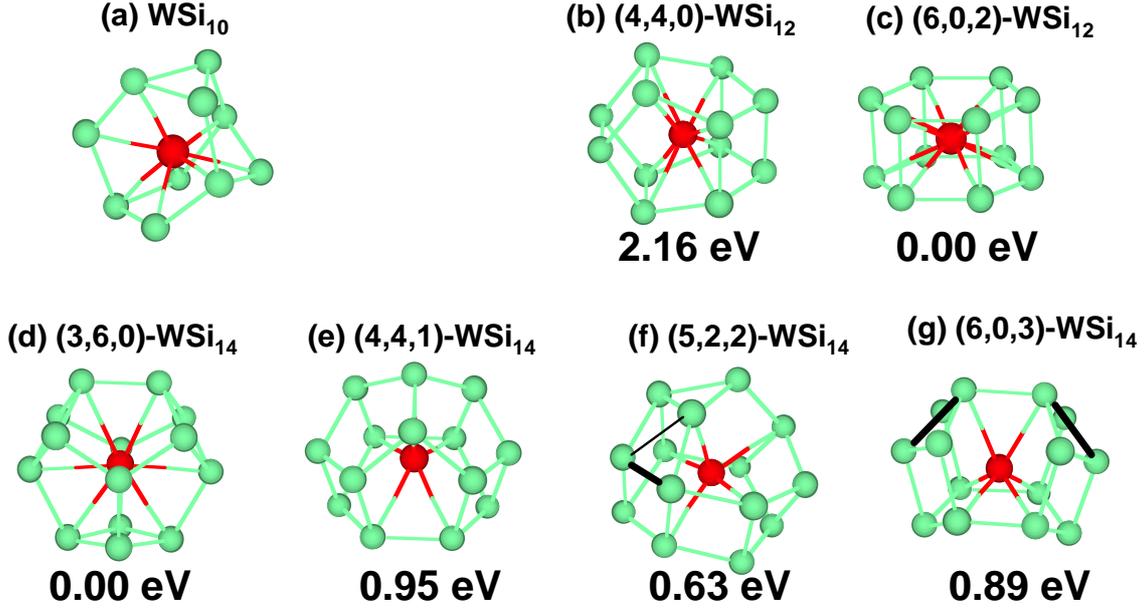

**Figure 1**. Optimized structures of WSi$_n$ clusters. Green and red balls represent Si and W, respectively. Green line segments between Si atoms are drawn when the inter-atom distances are less than 2.7 Å. The same applies to drawing line segments between W and Si atoms. In panels (f) and (g), thin (thick) black line segments should be supplemented (removed) to retrieve connectivity in the corresponding simple 3-polytopes. The bracketed three numbers in each panel represent ($f_4, f_5, f_6$). The energies of WSi$_{12}$ and WSi$_{14}$ are relative to those of panels (c) and (d), respectively. A stable structure of WSi$_{10}$ cluster (panel (a)) is not of a simple 3-polytope although Si atoms surround it.

**Table 1.** The number of the inner diagonals ($\delta_3$), the effective inner diagonals ($\delta_3^{eff}$) and relative energies ($\Delta E$) of WSi$_n$ ($n=12$ and $14$) clusters. The $f_k$ is the number of $k$-membered rings in the Si cage of a cluster. The $\delta_3^{eff}$ is calculated for various cut-off radii, (a) 1.0Å, (b) 0.6Å, (c) 0.3Å and (d) 0.25Å. The $\Delta E$ is measured relative to the total energy of the clusters with $(f_4,f_5,f_6)=(6,0,2)$ and $(3,6,0)$ for $n=12$ and $14$, respectively.

| $n$ | $(f_4,f_5,f_6)$ [structure] | $\delta_3$ | $\delta_3^{eff}$ | | | | $\Delta E$ (eV) |
|---|---|---|---|---|---|---|---|
| | | | (a) | (b) | (c) | (d) | |
| 12 | (4,4,0) [Fig.1(b)] | 20 | 8 | 4 | 0 | 0 | 2.16 |
| 12 | (6,0,2) [Fig.1(c)] | 18 | 6 | 6 | 6 | 6 | 0.00 |
| 14 | (3,6,0) [Fig.1(d)] | 34 | 4 | 4 | 4 | 4 | 0.00 |
| 14 | (4,4,1) [Fig.1(e)] | 33 | 11 | 5 | 1 | 0 | 0.95 |
| 14 | (5,2,2) [Fig.1(f)] | 32 | 13 | 4 | 2 | 2 | 0.63 |
| 14 | (6,0,3) [Fig.1(g)] | 31 | 9 | 7 | 4 | 0 | 0.89 |

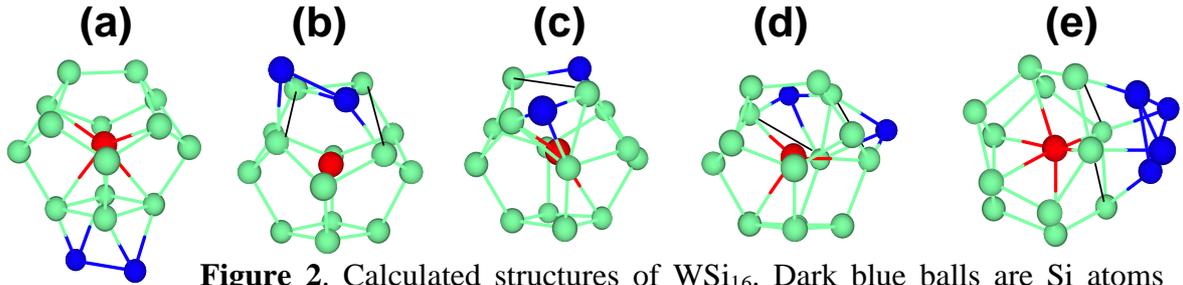

**Figure 2**. Calculated structures of WSi$_{16}$. Dark blue balls are Si atoms regarded "attached" to either the (3,6,0)-WSi$_{14}$ (panels (a), (b), (c) and (d)) or (6,0,2)-WSi$_{12}$ cluster (panel (e)). Other conventions are the same as in Fig.1. This result means that there is a maximum number of Si atoms to enclose a given metal atom with a single smooth Si cage.

In the present situation where there are much more Si atoms than a metal atom, fourfold coordination of each Si atom can be maintained by establishing one bond to the metal atom and remaining three to other three Si atoms. The topology of the network made by Si atoms then becomes that of a simple 3-polytope. As already suggested, the inner diagonals passing close to an encapsulated metal atom approximate the bonds between Si and metal atoms. Therefore the structure of the energetically favorable WSi$_n$ cluster has a Si cage of a simple 3-polytope with the maximum $\delta_3^{eff}$.

Another important result is that there appears a maximum cage size of WSi$_n$, beyond which the cage becomes unstable. In Fig.2, some of calculated structures of WSi$_{16}$ clusters are shown. The initial geometries were those of simple 3-polytopes with 16 vertices. Upon relaxation, all cages studied are completely distorted and become similar to either a (6,0,2) cage plus 2 Si atoms or a (3,6,0) cage plus 4 Si atoms.

In order to understand the existence of the maximum cage size, we may simply consider distances among the constituent atoms in addition to the above topological argument. Since the distances between Si atoms in the cage should not be sensitively changed depending on *n*, the size of the cage (or the volume of the hollow space inside the cage) may be larger for larger *n*. As a result, the average distance between the metal atom and the cage becomes larger. The position of the metal atom should be off-centered in the cage. This gives rise to regions in the cage where the Si atoms do not "see" the metal atom but interact themselves. Thus a cage with too large *n* should collapse into smaller cages, explaining why the threshold cage size exists.

*3.3   Electronic states of WSi$_n$*

The peculiar cage geometry of low-energy WSi$_{12}$ and WSi$_{14}$ clusters shown in Fig.1 is a result of a beautiful cooperation between the W-Si and Si-Si bonding interactions. The former is composed of *s-d* hybrid orbitals at the W atom and *s* and *p* orbitals of Si. Especially, the symmetry of the tungsten d orbitals favors the structure of the Si cage with a regular hexagonal prism of the lowest-energy WSi$_{12}$ cluster (Fig.1(c)) where the hybrid orbitals of W very efficiently overlap with those of the Si cage. The matching between the d orbitals of W with the Si cage in the lowest-energy WSi$_{14}$ cluster (Fig.1(d)) becomes less efficient, in which case the admixture of the W *s* orbital becomes stronger to make the angular distribution of the *d* orbitals more anisotropic about the W atom. Plotting all occupied molecular orbitals of the lowest-energy WSi$_{12}$ and WSi$_{14}$ clusters, one can clearly see twelve and fourteen hybrid W-Si bonding orbitals, respectively (not shown in the present paper). This result justifies assuming the inner diagonals of a cage passing close to the W atom to approximate the chemical bonds between W and Si atoms. This implies the possibility that one can *predict* promising candidates of cage structures of Si atoms around a metal atom at a given number of *n*: *configure n Si atoms in simple 3-polytopes with large $\delta_3$'s*. As for the Si-Si bonds, there are four, one of which is "*s*"-like and the other "*p*"-like, molecular orbitals composed of the Si *s* orbitals whose eigenenergies are all lower than those of the W-Si bonding orbitals. These states should contribute to a non-negligible part of total cage cohesion.

Here a question might arise as to the structure of WSi$_{12}$. Why is the Si cage of this cluster not of an icosahedron? If 12 Si atoms are arranged in an icosahedron, then $\delta_3=36$, much larger than $\delta_3=20$, the maximum possible value of 12-vertex simple 3-polytope (Fig.2(a)). In the icosahedral cage, the number of bonds between W and twelve Si atoms should be too large (or too many electrons are accommodated in the W-Si bonds) while each Si atom is fivefold coordinated with neighboring Si atoms, meaning that there are too few electrons to fill up the Si-Si bonds. Thus the icosahedral cage should be energetically unfavorable. In fact, we have found that the total energy of a WSi$_{12}$ cluster with a regular icosahedral Si cage is 6 eV higher than that of the (6,0,2) cage (Fig.1(c)). The HOMO of the icosahedral cluster is fivefold degenerated and occupied with only four electrons including spins, meaning strong instability against the distortion of the cage. The topology of the icosahedron belongs to what is called a *simplicial* 3-polytope [17]. Since each facet of any simplicial 3-polytope is a triangle, this topology is favored by elements such as boron, which creates doubly occupied three-centered bond in each facet.

An interesting aspect of the electronic structure of the lowest-energy WSi$_{12}$ cluster is the contrast in the distribution of its highest-occupied molecular orbital (HOMO) and lowest-unoccupied molecular orbital (LUMO), as shown in Fig.3. The former is *p*-like and distributed over the Si cage (Fig.3(a)) while the latter is *d*-like character strongly localized at the W atom. If one would be able to arrange the clusters in the regular positions, either in two or three dimensions, it might be possible to tailor the components of the *p*-bands (HOMO-originated) and the *d*-band counterparts (LUMO-originated) by tuning the spacing, orientation and the symmetry of the lattice points where the clusters are located.

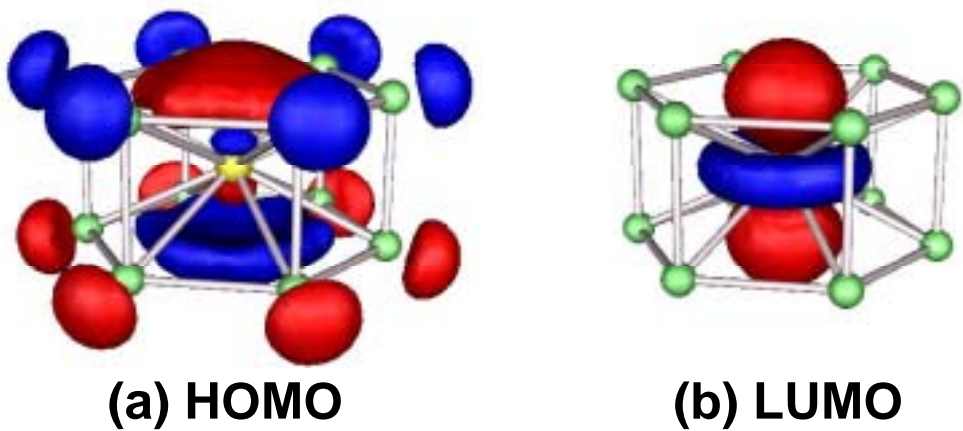

**(a) HOMO**          **(b) LUMO**

**Figure 3**. (a) Highest-occupied molecular orbital (HOMO) and (b) lowest-unoccupied molecular orbital (LUMO) of a WSi$_{12}$ cluster. Green and yellow balls represent Si and W atoms, respectively. The lobes colored with blue and red represent the wavefunctions with opposite phases.

## Conclusion

We have presented topology, energetics and electronic structures of WSi$_n$ clusters based on first-principles calculations. In a certain range of $n$ (=12 and 14), the Si atoms create fullerene-like cage, whose topology is of a simple $n$-vertex 3-polytope with a maximum number of the inner diagonals passing close to the W atom. The topological discussion developed here may have a potential impact to provide us with a more general and sophisticated definition of the fullerenes than those currently available.

## Acknowledgement

This work was partly supported by NEDO. This work has been conducted as a part of the Promoted Research Projects for High-Performance Computing by TACC-AIST.